\documentclass[copyright,creativecommons]{eptcs}
\providecommand{\event}{ICE 2016}

\usepackage{graphicx}
\usepackage{caption}
\usepackage{subcaption}
\usepackage{xcolor}
\hypersetup{%
        colorlinks=false,
        linkbordercolor=orange,
        pdfborderstyle={/S/U/W .4}
}

\title{Nominal Cellular Automata\thanks{Research partially supported by ISTI-CNR and by EU FP7 Project QUANTICOL, Grant agreement n. 600708.}}

\author{
Tommaso Bolognesi \qquad Vincenzo Ciancia
\institute{Istituto di Scienza e Tecnologie dell'Informazione ``A. Faedo''\\
	Consiglio Nazionale delle Ricerche \\ Pisa, Italy}
\email{\{tommaso.bolognesi, vincenzo.ciancia\}@isti.cnr.it}
}

\def\titlerunning{Nominal Cellular Automata}
\def\authorrunning{T. Bolognesi and V. Ciancia}

\begin{document}

\maketitle
\begin{abstract}
The emerging field of Nominal Computation Theory is concerned with the theory of Nominal Sets
and its applications to Computer Science. 
We investigate here the impact of nominal sets on the definition of Cellular Automata
and on their computational capabilities, with a special focus on the emergent behavioural properties
of this new model and their significance in the context of computation-oriented interpretations of physical phenomena.
A preliminary investigation of the relations between Nominal Cellular Automata and
Wolfram's Elementary Cellular Automata is also carried out.
\end{abstract}

\section{Introduction}
\label{sect:Introduction}
%

\emph{Nominal sets} are a modern recollection of the \emph{permutation} model that Fraenkel and Mostowski had used to prove independence of the \emph{Axiom of Choice} from the other axioms of set theory, and was re-introduced in Computer Science as the grounds for modelling \emph{abstract syntax} in the presence of \emph{binding} \cite{GabbayP99}. The theory of nominal sets has given rise in the last decades to a novel computing paradigm, based on the notion of \emph{pure name}, with proven capacity to embrace and enhance the full spectrum of Theoretical Computer Science, including syntax and semantics \cite{GabbayP99,MontanariP00}, automata theory \cite{BojanczykKL11}, regular expressions \cite{KozenMP015,DGabbayC11}, and several other foundational topics, including the most famous theoretical device, namely \emph{Turing machines} \cite{BojanczykKLT13}.

Cellular automata \cite{ref:Ilachinski2001} are a model of computation originally introduced in the 1940's by Stanislaw Ulam and John von Neumann
for modelling biological self-reproduction. 
Cells are arranged in regular, finite-dimensional grids, and  assume values from a finite alphabet.
Cell updatings occur simultaneously at discrete time instants (principle of \emph{synchrony}), the new value of each cell depending 
of the values of the cells in its neighborhood (principle of \emph{locality}).

ECA's (for Elementary Cellular Automata) \cite{ref:nks} are the simplest form of CA: 
cells are arranged in a 1D, infinite, or finite and circular, array, they assume binary values,
and cell neighbourhoods have \emph{diameter} 3, meaning that the evolution of each cell is determined by its
immediate left and right neighbours.  In other words, 
the value $c_{i}(t+1)$ of cell $c_{i}$ at time $t+1$ depends on the value at time $t$ of $c_{i}$ itself 
and of its immediate left and right neighbors,
and is expressed by $f(c_{i-1}(t), c_{i}(t), c_{i+1}(t))$, where $f$ is a boolean function 
used for all cells (principle of \emph{uniformity}).
There exist $2^{8} = 256$ boolean functions of three variables, thus there exist 256 distinct ECA.
According to Wolfram's numbering scheme, if $f(1,1,1) = b_{1}, f(1, 1, 0) = b_{2}, ..., f(0, 0, 0) = b_{8}$,
then the ECA based on function $f$ is numbered by the decimal representation of bit tuple $(b_{1}, ..., b_{8})$.

Being based on principles of locality and uniformity, CA's have been used for modelling a wide array of natural phenomena, 
from crystal formation to fluid dynamics,
from reaction-diffusion in chemistry to pattern formation in biologic substrata. 
The contrast between the extreme simplicity of their definition and the richness of their emergent behaviours,
as widely explored in \cite{ref:Wolfram84, ref:nks}, has induced some scientists to attribute to CA's, 
or similar simple models,
a fundamental role for explaining the complexity of the physical universe as a whole.
The idea that the complexity observed in Nature be the manifestation of the emergent properties of
a computation taking place at the tiniest spacetime scales is sometimes referred to as the
`computational universe conjecture'.

The quest for simple theoretical models of digital space­time is still open. Classical cellular automata provide a foundational example. However, CA already assume a universal meaning of colours in cells. In this work, instead, we consider a model which is solely based on equality, where the meaning of colours is only defined at the local level. In order to achieve this, we resort to the machinery of \emph{nominal sets}. The main purposes of this work are 
(i) to come up with a revised definition of cellular automata based on nominal computing concepts, and
(ii) to present a \emph{preliminary} exploration and discussion of the potential advantages 
that Nominal Cellular Automata might offer over classical CA's
when dealing with the computational universe conjecture and, 
more generally and abstractly, with the emergence of complexity from simple premises.

In Section \ref{sect:DefsAndCounts} we provide the general definition of Nominal Cellular Automaton,
which merges naturally the classical definition of Cellular Automata (CA) with the basic assumptions of
nominal computing, and we classify the space of NCAs, determining class sizes.

In Section \ref{sect:ENCA} we examine the simplest form of NCA, that we call ENCA (`E' for `Elementary').

In Section \ref{sect:DirectSimulation} we start studying the relations between ENCA and Wolfram's Elementary CA (ECA), 
identifying the subset of the ECA's that can be \emph{directly} simulated by ENCA's.

In Section \ref{sect:Simulation} we introduce a simple technique for simulating any ECA - including Turing-complete ECA 110 - by NCA's,
thus obtaining a first (but still cumbersome) instance of  a Turing-complete NCA.

In Section \ref{sect:Behaviors} we present some of the emergent patterns that can be observed in the spatio-temporal diagrams of NCA's,
compare them with the classical CA scenario, and speculate on the potential benefits that the new model might offer with respect to 
the computational universe conjecture.
%
%
\section{Nominal Cellular Automata (NCA): definitions and classes}
\label{sect:DefsAndCounts}
%
%
While in traditional cellular automata the values assumed by cells range in a \emph{finite} alphabet of symbols, or states
(sometimes called `colours'),
in a \emph{Nominal Cellular Automaton} (NCA) cells are drawn from a \emph{countably infinite} alphabet: 
their values range in an infinite set of symbols, or \emph{names}, that we denote $N$
and conveniently represent by the natural numbers.  
A name denotes nothing but itself, and the only operations on names are: 
(i) compare names for equality,
(ii) make a copy of a name,
(iii) create a brand new name.

Similar to classical CA, the \emph{transition rule} of a NCA establishes the next value of cell $c_{i}$
based on the current values of the cell and of those in its neighbourhood or, more precisely,
on the \emph{equality pattern} exhibited by these names.
We restrict here to one-dimensional CA's, in which the cells are arranged in a one-dimensional finite or infinite array.
We shall use the term \emph{context} of cell $c_{i}$ to denote the tuple consisting of $c_{i}$ and some of its left and right neighbours, 
and the term \emph{diameter} to precisely define the number of cells in the context (including $c_{i}$).
For example, the diameter-3 context of $c_{i}$ is  $(c_{i-1}, c_{i}, c_{i+1})$.

The new value to be assigned to a cell by the transition rule 
can be either one of the names found in the context at a given position, or a new name.
The idea is that the rule is not allowed to access the names generated up to the current computation step:
it can only access those immediately found in the context to which it applies, or create a brand new one.

Let us count the number of possible transition rules, thus NCA instances, for context diameter $d$.

The transition rule has a different component for each different context. 
Thus we first need to determine the number of distinct contexts, or, more precisely, 
distinct \emph{equality patterns} that are possible for that diameter
- a number that we denote $X(d)$.
The \emph{equality pattern} of context $c = (c_{1}, ..., c_{d})$ is the information that indicates which element of the context is equal to which other element,
with elements identified by their position in $c$.  
Formally, we could define it as the equivalence class of the $d$-tuples $b = (b_{1}, ..., b_{d})$ over $N$ such that 
$b_{i} = b_{j}$ if and only if $c_{i} = c_{j}$, with $i, j \in \{1, 2, ..., d\}$.
For example, the equality pattern for context (38, 4, 4, 7, 11, 7)
is a countably infinite set of 6-tuples including, among others, 
(2, 4, 4, 1, 0, 1), (0, 1, 1, 2, 3, 2), and of course (38, 4, 4, 7, 11, 7) itself.

Then, we can easily establish that $X(d)$ is the same as the number of partitions of a \emph{set} of $d$ different objects,
which we can represent by the first $d$ integers: $\{1, 2, ..., d\}$.
The reason is that we can establish an obvious one-to-one correspondence between the equality pattern and the partition.
For example, the equality pattern for context (38, 4, 4, 7, 11, 7) corresponds to partition $\{\{1\}, \{2, 3\}, \{4, 6\}, \{5\}\}$:
the latter neatly indicates that the first and fifth element of the context are different from each other and from all other elements,
while the elements at positions 2 and 3 are equal, but different from all other elements in the context; 
the same holds for the elements at positions 4 and 6.  
It is also clear that the partition identifies all and only the tuples of the equality pattern.

The number of partitions of a set of $d$ different elements - thus our $X(d)$ - is known as \emph{Bell number},
or \emph{exponential number}, and is denoted $Bell(d)$.  
The first Bell numbers, starting with $d = 1$, are 1, 2, 5, 15, 52, 203, 877, 4140, 21147, 115975, 678570.
One way to generate these numbers
\footnote{Further information at
http://mathworld.wolfram.com/BellNumber.html
and
https://oeis.org/A000110.
}
is to use the recurrence:

\[
Bell(n) = \sum_{k=0}^{n-1}Bell(k) {n-1 \choose k}.
\]

For each context involving cell $c_{i}$ and its neighbours
the transition prescribes a reaction, i.e. it defines the name to be assigned to $c_{i}$ next, which can be
one of the names found in the context, or
a brand new name.
 Note that we use the term `context' for denoting both an \emph{actual tuple} of names 
 and the \emph{class of tuples} that share the same equality pattern;
 strictly, the nominal `viewpoint' only understands the latter. 
  
 Let $\cal{C}$$^{d}$ = $\{C^{d}_{1}, C^{d}_{2}, ..., C^{d}_{Bell(d)}\}$ be the set of contexts of length $d$ and let $\alpha(C^{d}_{i})$ be the
 number of distinct names featured by context $C^{d}_{i}$.  
 The number of possible reactions relative to context $C^{d}_{i}$ is $\alpha(C^{d}_{i}) + 1$,
 where $\alpha(C^{d}_{i})$ counts the options for copying and re-using a name found in the context
 and +1 reflects the creation of a new name.
 We can then express the number of distinct instances of diameter-$d$ NCA's as follows:
 \begin{equation}
SpaceSize(d) = \prod_{i=1}^{Bell(d)} (\alpha(C^{d}_{i})+1).
\label{eq:SpaceSizes}
\end{equation}
We have implemented a simple context enumeration scheme and algorithmically computed $SpaceSize(d)$
for various values of $d$.  These are shown in the table below.



\begin{table}[ht]
\begin{center}
\begin{footnotesize}
 \begin{tabular}{| c | c |} 
 \hline
  \emph{d} & \emph{SpaceSize} \\ 
 \hline \hline
 2 & 6 \\
 \hline 
 3 & 216 \\
 \hline 
 4 & 89579520 \\
 \hline 
 5 & 1893214811085172899840000000000 \\
  \hline
 6 & $\approx 10^{127}$ \\
 \hline
 7 & $\approx 10^{584}$ \\
  \hline
 8 & $\approx 10^{2899}$ \\
 \hline
 \end{tabular}
\caption{
{\footnotesize Sizes of NCA spaces for various values of context diameter \emph{d}.}
}
\label{tab:spaceSizes}
\end{footnotesize}
\end{center}
\end{table}

\begin{figure}[h]
\centering
\includegraphics[scale=.6]{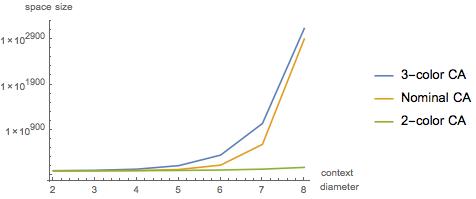}
\caption{{\footnotesize  The space of NCA's approximates in size that of 3-color CA's, for context sizes $\leq 8$.}}
\label{fig:spaceSizes}     
\end{figure}

In Figure \ref{fig:spaceSizes} we compare these space sizes with those of 2-color and 3-color CAs with the same range of context diameters.
\footnote{An `\emph{n}-color' CA is one in which cells are $n$-state automata.}
%
%
%
%
\section{Elementary NCA (ENCA)}
\label{sect:ENCA}
%
%
In analogy with ECA (Wolfram's Elementary Cellular Automata), we define
ENCA (Elementary Nominal Cellular Automata) as the subclass of NCA with context diameter 3.

The number of different contexts of size 3 is small: $Bell(3) = 5$.
Thus, we list all of them in Table \ref{tab:EncaRuleCount}, 
and identify for each context $C$ the possible reactions and their number, which is $\alpha(C)+1$.
An ENCA behaviour is defined by indicating the reaction for each possible context. 
It follows that the ENCA space size is given by the product of the numbers in the rightmost column of the table, namely 216:
the ENCA's are even less than the ECA's (256).

\begin{table}[ht]
\begin{center}
\begin{footnotesize}
 \begin{tabular}{| c | c | c | c |} 
 \hline
 \emph{i} & \emph{context} $c_{i}$ & \emph{possible reactions} & $\alpha(c_{i}) + 1$ \\ 
 \hline \hline
  1 & (a, a, a) & a $|$ n & 2 \\
 \hline 
 2 & (a, a, b) & a $|$ b $|$ n & 3 \\
  \hline 
 3 & (a, b, a) & a $|$ b $|$ n & 3 \\
  \hline 
4 & (a, b, b) & a $|$ b $|$ n & 3 \\
  \hline 
 5 & (a, b, c) & a $|$ b $|$ c $|$ n & 4 \\
 \hline
 \end{tabular}
\caption{
{\footnotesize Possible reactions for each context of length 3.  Letter `n' denotes a brand new name.
The product of the numbers in the third column yields the ENCA space size 216.}}
\label{tab:EncaRuleCount}
\end{footnotesize}
\end{center}
\end{table}

We can conveniently use a \emph{mixed basis} (or \emph{mixed radix}) numeral system for associating a number to each ENCA,
using bases $\beta(i) = \alpha(c_{i}) + 1$, $i = 1, ..., 5$, i.e.\ (2, 3, 3, 3, 4).
For example, the mixed basis notation:
\[
0_{2}2_{3}1_{3}0_{3}3_{4}
\]
identifies number $3 + 0(4) + 1(3*4) + 2(3*3*4) + 0(3*3*3*4) = 87$, where the addends are obtained by scanning the above
expression from right to left, and by using the product of the elements of the growing suffixes of tuple (2, 3, 3, 3, 4).
The smallest number expressible with this notation is $0_{2}0_{3}0_{3}0_{3}0_{4} = 0$, while
the largest number is $1_{2}2_{3}2_{3}2_{3}3_{4} = 215$, 
establishing the range [0, 215] of indices for the 216 ENCA's. 

The transition rule of ENCA 87 is directly determined by the above mixed basis notation for number 87.
The digits (0, 2, 1, 0, 3) are associated to the contexts listed from top to bottom in Table \ref{tab:EncaRuleCount}:
the \emph{first} digit, 0, indicates that, if the \emph{first} context $(a, a, a)$ is detected, 
then the first reaction - $a$ - is taken, and name $a$ is assigned to the central cell 
(we number the alternative items at each entry of column `\emph{possible reactions}' from left to right, starting from 0). 
The \emph{second} digit, 2, indicates that, if the \emph{second} context $(a, a, b)$ is detected, then the third possible reaction - $n$ - is taken,
and a brand new name is assigned to the central cell; and so on.

We shall consider \emph{circular} arrays of length $n$, where cells $c_{n}$ and $c_{1}$ are regarded as neighbours.
A technique widely exploited in \cite{ref:nks} for a preliminary assessment of the computational capabilities of 
CA's consists in visually inspecting their diagrams when the computations start from uniform or random initial conditions.
In these diagrams we shall represent different names by different grey levels.

\begin{figure}[h]
\centering
\includegraphics[scale=.6]{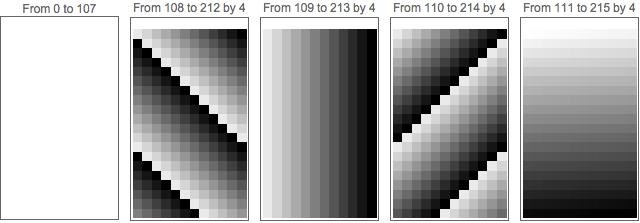}
\caption{{\footnotesize The five types of ENCA behaviour starting from uniform initial conditions - same name in all cells.
The labels identify the ENCA rule numbers that correspond to each behaviour.}}
\label{fig:encaTypesFromUniformInit}     
\end{figure}

The simplest uniform initial condition consists of an array of equal names.
Figure \ref{fig:encaTypesFromUniformInit} shows the only five types of behaviour that are obtained in this case,
and indicates the subset of ENCA corresponding to each of them.
%
%
In the second, third and fourth case, $n$ new names are introduced at the first step, where $n$ is the number
of cells in the circular array (in Figure \ref{fig:encaTypesFromUniformInit} we set $n = 12$), 
and then 
shifted to the right (rules 108 to 212, by 4), 
not shifted (rules 109 to 213, by 4), or
shifted to the left (rules 110 to 214, by 4).
Only in the fifth case $n$ new names keep being created at each step, undefinitely.

\begin{figure}[h]
\centering
\includegraphics[scale=.42]{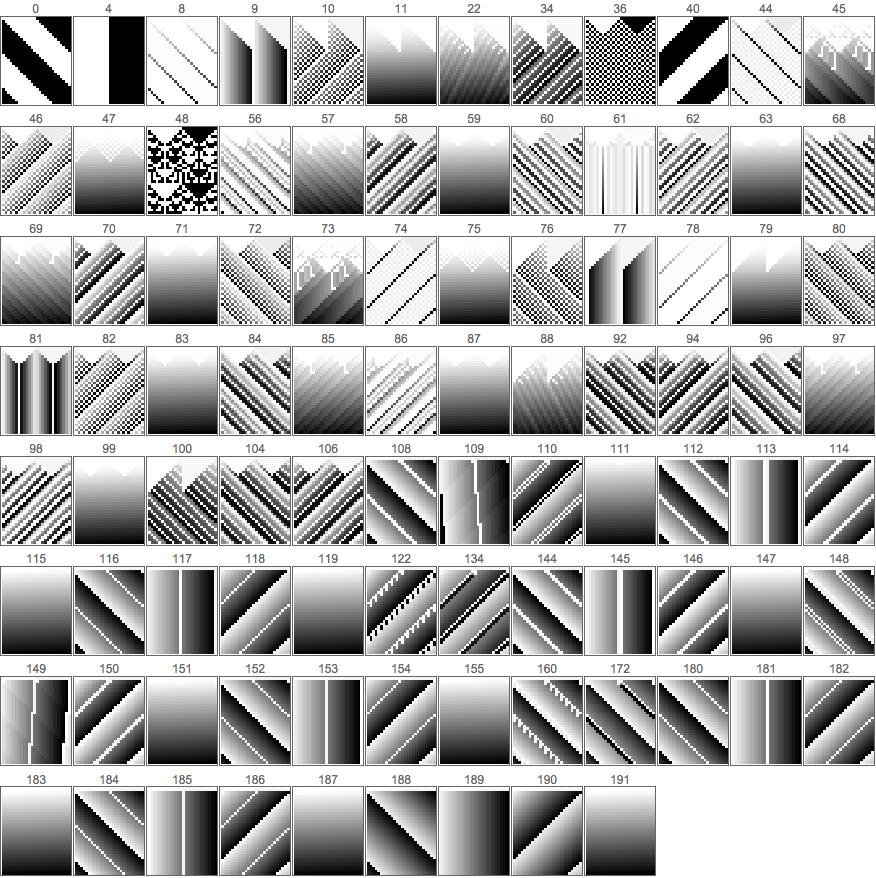}
\caption{{\footnotesize The 93 types of ENCA behaviour starting from two adjacent runs of thirteen 0's and thirteen 1's.
The labels identify the smallest ENCA rule that corresponds to each behaviour.}}
\label{fig:encaTypesFromSemiUniformInit}     
\end{figure}

A richer variety of behaviours is obtained by using \emph{two} runs of equal names in the initial condition -
a tuple of 0's followed by a tuple of 1's.
Figure \ref{fig:encaTypesFromSemiUniformInit} shows the 93 types of behaviours obtained, each labelled by
the number of the smallest rule that realizes it.
%
%

With an initial cell array in which all names are different, only four distinct behaviours are observed.
Essentially, these are already covered in Figure \ref{fig:encaTypesFromUniformInit}, where the
the second row in each of the four rightmost diagrams is a tuple of distinct names
(this fact is not apparent in the fifth diagram of that figure, due to the presence of a high number of different names/grey levels). 

The most interesting cases are those in which the initial condition contains fewer distinct names than its size, with some duplicates.
We postpone the illustration of some of these cases to the closing section of the paper.
%
%
\section{Direct simulation of ECAs by ENCAs}
\label{sect:DirectSimulation}
%
%
What are the relations between ENCA's and ECA's?
We start answering this question by identifying a subset of ECA's that can be \emph{directly} simulated by ENCA's.
By `directly' we mean that no pre-processing of the initial ECA configuration should be necessary
for the ENCA to faithfully reproduce the ECA computation, bit by bit.

Let us consider the conditions an ECA rule must satisfy for an ENCA to be able to directly simulate it.
Similar to ENCA rule numbers, an ECA rule number, expressed in binary form (thus in \emph{fixed} basis), 
identifies the reactions associated to each of the 8 possible contexts - bit triples.
Consider, for example, ECA 142.  The bit octet $(b_{1}, b_{2}, ..., b_{8})$ representing 142 is (1, 0, 0, 0, 1, 1, 1, 0):
these bits represent the new values of the central cells of, respectively, the eight contexts ((1, 1, 1), (1, 1, 0), (1, 0, 1), (1, 0, 0), (0, 1, 1), (0, 1, 0), (0, 0, 1), (0, 0, 0)).
For an ENCA, contexts $(b_{1}, b_{2}, b_{3})$ and $(b_{1}', b_{2}', b_{3}')$, where the prime symbol denotes bit flipping, are indistinguishable,
thus the reaction is the same, and will be to copy one of the distinct names found at a fixed position in the corresponding triplet.
This implies that the simulated ECA must react with flipped bits to flipped bit triples.
Furthermore an ENCA can only copy names found in the inspected context, hence the reaction to (1, 1, 1) can only be `1',
and the reaction to (0, 0, 0) must be `0' (as also implied by the previous observation).
It follows that an ECA rule can be directly simulated by an ENCA only if its binary octet is of form $(1, b_{2}, b_{3}, b_{4}, b_{4}', b_{3}', b_{2}', 0)$.
Note that the octet above for ECA rule 142 satisfies these conditions.
Three independent bits are sufficient to identify octets of this form. 
In conclusion, there are 8 ECAs that can be directly simulated by ENCA's, namely 142, 150, 170, 178, 204, 212, 232, 240,
since these are the only numbers in the range [0, 255] whose binary representation satisfies
the above symmetry conditions.

For identifying the ENCA's that simulate ECA $x$, with $x$ ranging in the above set of 8 elements,
we must consider the binary representation $(b_{1}, b_{2}, ..., b_{8})$ of $x$
and use the bit tuple $(b_{8}, b_{7}, b_{6}, b_{5})$, where $b_{8}$ is necessarily 0.
The bits $(b_{8}, b_{7}, b_{6}, b_{5})$ correspond to choices to be made by the ENCA rules within 
the set of `\emph{possible reactions}' of Table \ref{tab:EncaRuleCount}.
For example, for ECA 142 we have $(b_{8}, b_{7}, b_{6}, b_{5}) = (0, 1, 1, 1)$.  
This means that an ENCA that directly simulates this ECA must implement the following reactions
to the first four contexts listed in Table \ref{tab:EncaRuleCount}:
$(a, a, a) \rightarrow a$,
$(a, a, b)  \rightarrow b$,
$(a, b, a) \rightarrow b$, 
$(a, b, b)  \rightarrow b$.
The reaction to context $(a, b, c)$ can be expressed as a base-4 digit $z$ that selects in $a | b | c | n$,
and is irrelevant since this context never occurs in ECA computations:
this explains why 4 distinct ENCA's simulate the same ECA.
The codes of the latter are obtained by interpreting the five digits $(b_{8}, b_{7}, b_{6}, b_{5}, z)$ in the mixed basis notation
introduced earlier for numbering ENCA's.

\begin{figure}[h]
\centering
\includegraphics[scale=.56]{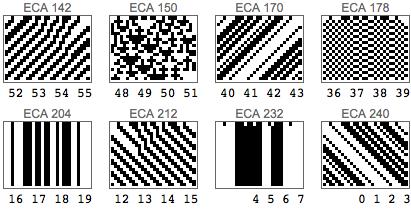}
\caption{{\footnotesize The eight ECA's that can be \emph{directly} simulated, (their corresponding ENCA is exactly the same)}}
\label{fig:fourENCAsForEachECA}     
\end{figure}

In Figure \ref{fig:fourENCAsForEachECA} we show the eight ECA's that can be directly simulated,
and the corresponding ENCA's.  The computations start with random bit configurations.
%
%
%
%
\section{Simulation of ECAs by NCAs}
\label{sect:Simulation}
%
%
By appropriately coding the initial conditions, we can simulate \emph{any} of the 256 ECA's 
by using NCA's of sufficiently large diameter.

We show here how to simulate a generic ECA $x$ with a diameter-12 ENCA.

Let $f:\{0, 1\}^3 \rightarrow \{0, 1\}$ be the boolean function that yields the next value of a cell 
based on the current value of its context $c$, for ECA $x$,
and let $F:N^{12} \rightarrow N$ ($N$ are the non-negative integers) be the (equivariant) function that yields the next value of a cell 
based on the equality pattern of the current context $C$, for the ECA $X$ designed to simulate $x$.

The coding of the initial configuration  of $x$ consists in inserting the three names (0, 1, $p$),
with $p$ brand new, after each bit $b_{i}$, so that:
\[
c = (b_{1}, b_{2}, b_{3}, b_{4} \dots) \rightarrow C = (b_{1}, 0, 1, p, b_{2}, 0, 1, q, b_{3}, 0, 1, r, b_{4}, 0, 1, s \dots).
\]
The diameter-12 ENCA $X$ that simulates ECA $x$ detects the position of the triple of mutually distinct names,
say $p$, $q$, $r$, within the current context $C$, with $q$ and $r$ at respective distances 4 and 8 from $p$ - and reacts accordingly.  
There are four possible cases, that we list below with their associated actions.
\begin{description}
\item[$p$, $q$, $r$ at $C_{4}$, $C_{8}$, $C_{12}$.] 
Then we know that $C_{1}, C_{5}, C_{9}$ are the original ECA bits and set $F(\mbox{C}) = f(C_{1}, C_{5}, C_{9})$.
\item[$p$, $q$, $r$ at $C_{3}$, $C_{7}$, $C_{11}$.] 
Then we know that $C_{1}, C_{5}, C_{9}$ are 0's and we copy one of them, setting $F(\mbox{C}) = C_{1}$.
\item[$p$, $q$, $r$ at $C_{2}$, $C_{6}$, $C_{10}$.] 
Then we know that $C_{1}, C_{5}, C_{9}$ are 1's and we copy one of them, setting $F(\mbox{C}) = C_{1}$.
\item[$p$, $q$, $r$ at $C_{1}$, $C_{5}$, $C_{9}$.] 
Then we set $F(\mbox{C}) = n$, where $n$ is a brand new name.
\end{description}

The idea is that function $F$ mimics function $f$ in one case, and guarantees to produce, respectively, 0's, 1's and new names
in the other three cases, so that the possible patterns of diameter-12 contexts are replicated at each new row.

We have implemented the above rule in \emph{Mathematica}, and verified that the computation of the context-12 NCA, once decoded
by retaining only the names at columns 1, 5 and 9, reproduces the computation of the chosen ECA $x$.

 %
\subsection{Turing-universality: simulating ECA 110}
\label{subsect:Simulating110}
%
ECA 110 is Turing-universal \cite{ref:Cook2004}, thus any device able to simulate this elementary
CA inherits its universality.  In light of the construction above, we conclude that diameter-12 NCA's include at least one
universal instance.

In fact, we can do slightly better, and find a diameter-9 NCA  - call it $Z$ - that simulates ECA 110.
In this case, the coding consists in inserting only \emph{two} names - (0, $p$),
where $p$ is a new name - after each bit $b_{i}$ of the initial ECA 110 configuration $c$:
\[
c = (b_{1}, b_{2}, b_{3} \dots) \rightarrow C = (b_{1}, 0, p, b_{2}, 0, q, b_{3}, 0,  r  \dots).
\]

$Z$ is essentially built as described for the diameter-12 case, except that now only three cases are distinguished,
relative to the position of the `trident' of distinct names $p$, $q$, $r$.

The possibility of reducing to diameter 9 is a consequence of the lucky circumstance 
that in ECA 110 (01101110 in binary) function $f$ is such that $f(1,1,1) = 0$ and $f(0,0,0) = 0$
for the only two bit triplets that involve just one name.  
In both cases a 0 must be created, and in both cases (that a nominal CA cannot distinguish) function $F$ `blindly'
finds this 0 in a predefined position of the inspected context, in virtue of the additional 0's inserted by the coding.

\begin{figure}
\centering
\includegraphics[height=4cm]{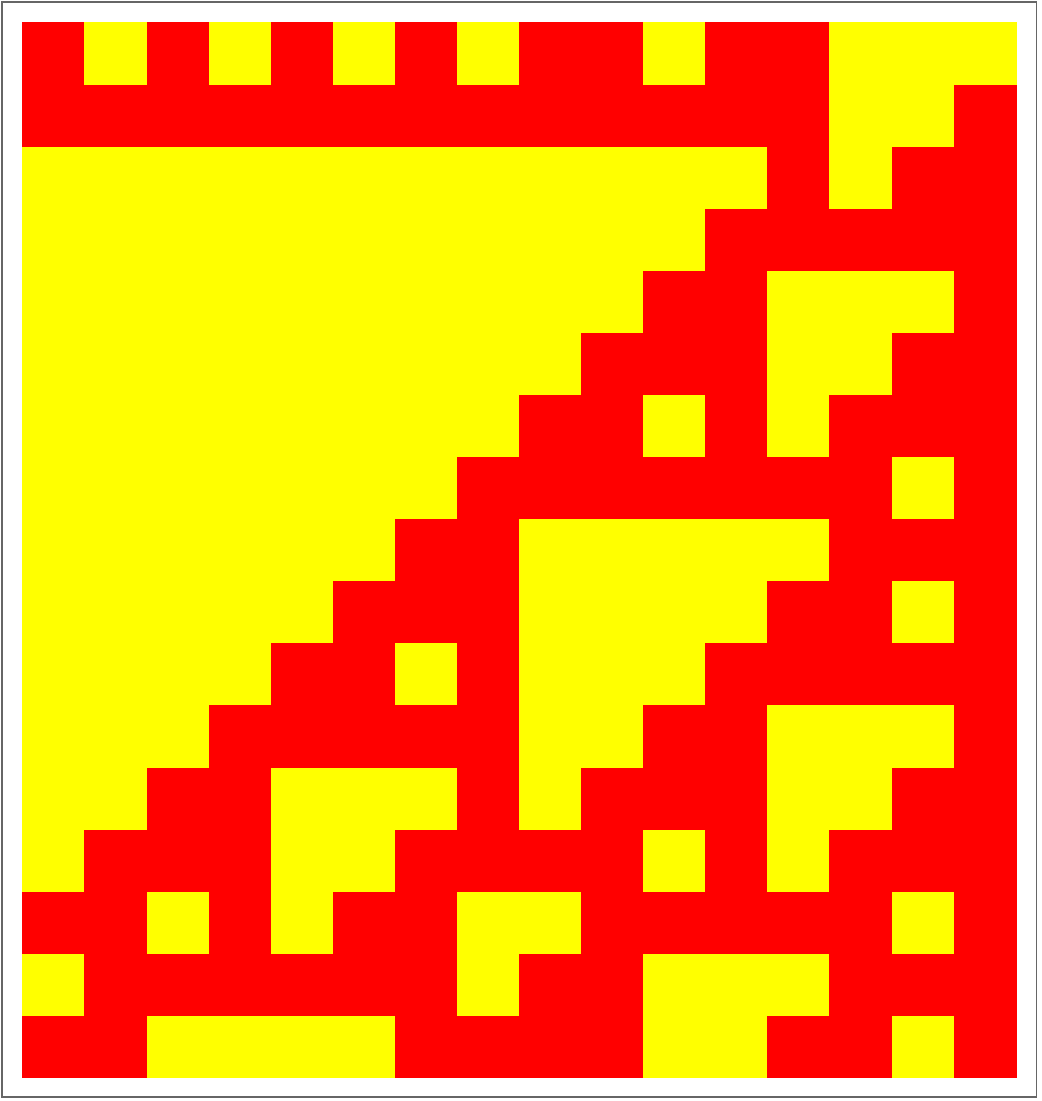} \hskip 20pt \includegraphics[height=4cm]{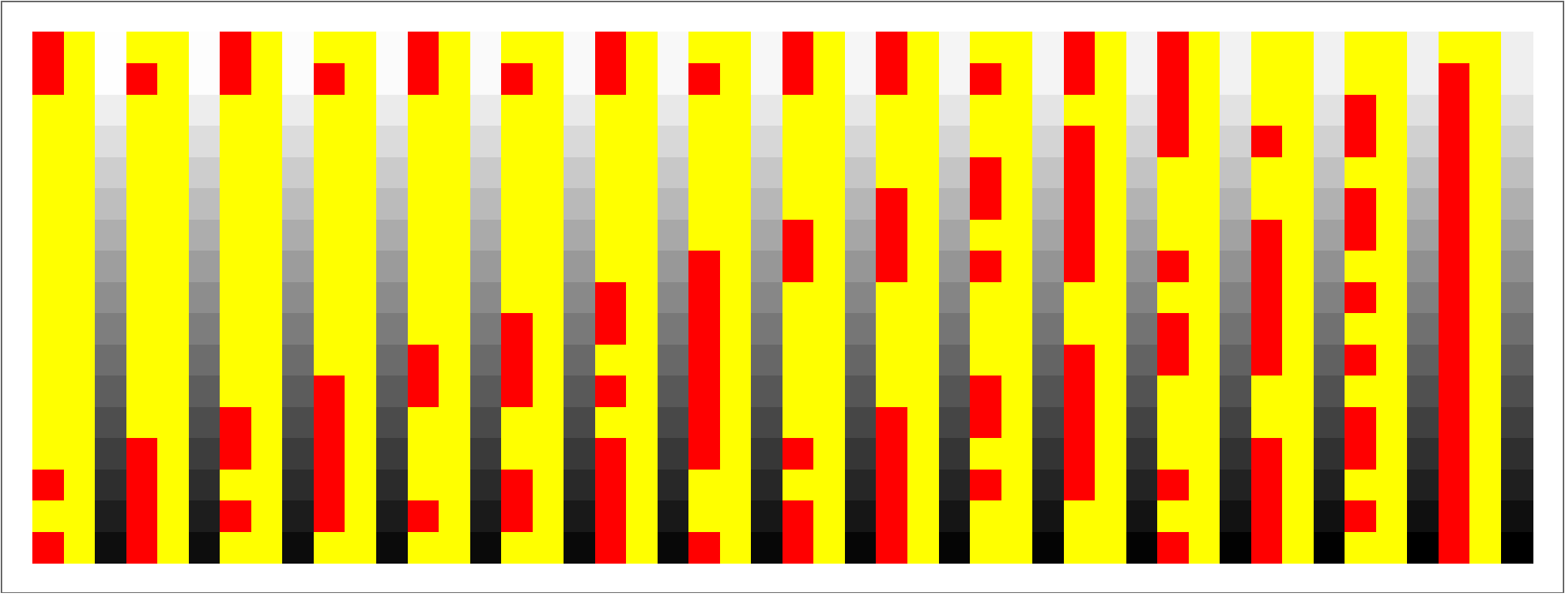}
 \caption{\label{fig:NCA110}An example ECA 110 computation and the corresponding computation in the simulating NCA.}
\end{figure}

In \autoref{fig:NCA110} we show an example ECA 110 computation and the corresponding NCA.

%
%
\section{Emergent behaviours}
\label{sect:Behaviors}
%
%
How does the adoption of the nominal viewpoint impact on the emergent properties exhibited by cellular automata?


One of the most spectacular emergent features of CA's is that of \emph{digital particles} (we shall drop the attribute `digital' in the sequel).
First observed by Zuse in \cite{ref:Zuse1970}, particles became popular with the `gliders' and various forms of `spaceships' arising in Conway's Game of Life,
a two-dimensional CA, and were subsequently discovered also by Wolfram in ECA's.
A particle is a periodic structure that moves across the spacetime CA diagram,
where space and time are associated, respectively, to the horizontal and the vertical dimension.
When started from a random array of bits, a handful of different particles invariably emerge in ECA 110 \cite{ref:nks},
that draw linear trajectories against a periodic background,
moving slower than `the speed of light' -- one horizontal cell per vertical step -- 
and engaging in various interactions. 
These particles play a crucial role for proving the Turing completeness of ECA 110 \cite{ref:Cook2004}.


The adoption of the nominal viewpoint has an impact on the way we may define particles.

In light of the unbounded name creation feature, 
it seems natural to track the moves of an individual name $p$ across spacetime
and consider it as a particle trajectory.  One may then study the interactions with the trajectory of some other name $q$.
In this case, the \emph{shape} of the trajectory (ignoring cell contents) does not have to be regular -- periodic --,
and it may even branch.

On the other hand, one might look for structures of bounded spatial width that move across spacetime,
formed by a \emph{finite} set of names that repeatedly occur, virtually forever, insisting that the shape of the trajectory be also periodic.
This is the case of the well known ECA 110 particles.
These structures are easily detected, visually, in our grey-level diagrams, since the periodic pattern of the names
is directly reflected in that of the grey levels.

Although the availability of an unbounded supply of names represents a departure from classical CA's
in that it potentially yields an infinite set of distinct particles, 
we still wish to call both of the mentioned types of structure `classical particles', 
since their identity is associated with one name, or a finite set of them, as in classical CA's.
 
\begin{figure}[h]
\centering
\includegraphics[scale=.56]{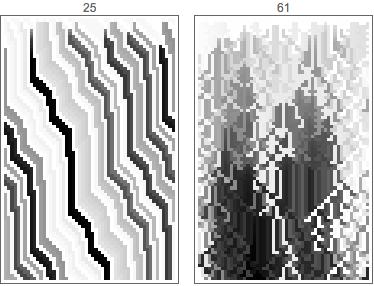}
\caption{{\footnotesize Classical particles in ENCA's 25 and 61.}}
\label{fig:particlesIn25and61}     
\end{figure}

Figure \ref{fig:particlesIn25and61} shows computations of ENCA's 25 and 61 from random bit tuples,
that lend themselves to the identification of examples of `classical' particles.  In ENCA 61 particles assume a tree-like form,
since some name gets duplicated at some step. 

 %
%

But the nominal viewpoint offers more: a particle can also be conceived as a spacetime substructure of limited width
in which periodicity holds \emph{only} in nominal sense, i.e. only at the level of the equality pattern.
It is clear that this implies the continuous appearance of new names along the trajectory, 
breaking `classical' periodicity.
\footnote{Note that a classical particle with periodic trajectory shape is periodic \emph{also} in nominal sense: 
the equality pattern is periodic too, that is, it is unaffected by a name permutation:
the crucial difference between classical and nominal particle is that the latter may involve an infinite set of names.}

\begin{figure}[h]
\centering
\includegraphics[scale=.56]{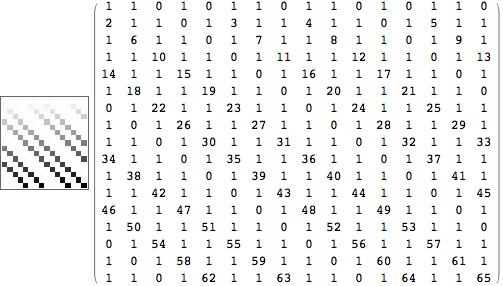}
\caption{{\footnotesize Classical and nominal particles in ENCA 8.}}
\label{fig:enca8particles}     
\end{figure}

For example, when started from a uniform random distribution of bits, ENCA 8 is the first member of the family to trigger unbounded name creation.
Figure \ref{fig:enca8particles} illustrates its behavior, showing both actual names (at the right) and their grey level representation (at the left).  
In this case, particularly simple diagonal particles of both types discussed above are easily detected.
%
%
%
For example, the diagonal formed by instances of name `1' , starting at matrix cell (1,1) is a classical particle,
while the line immediately below, starting at matrix cell (2,1) with names 2, 6, 10, 15, 19..., represents a properly nominal particle,
involving also mutually distinct names -- in fact, \emph{only} distinct names, in this case.
Incidentally, note that the numeric pattern is not regular -- the numbers along the trajectory do not form an arithmetic progression --
ruling out the the most obvious conceivable techniques for the automatic detection of these structures. 
(On the contrary, ENCA 28 produces particles of both types that do not move in space, but only in time, 
and the names in the nominal particles are in arithmetic progression.)

Started from random bits, ENCA 11 is the first automaton to exhibit `maximum creativity': 
it quickly settles into a steady state in which new names are generated for each cell at each step,
thus yielding a dynamics that appears analogous to the total uniformity observed in the simplest ECA's.
In principle, any orderly arrangement of cells that cuts across this spacetime diagram at some angle
could be regarded as a nominal particle, in this case!

\begin{figure}[h]
\centering
\includegraphics[scale=.56]{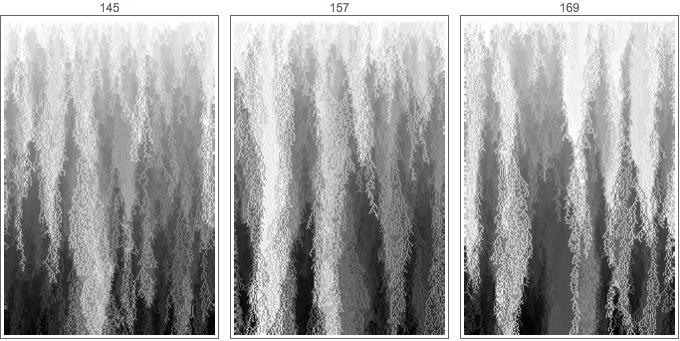}
\caption{{\footnotesize ENCA's 145, 157, 169, from the same initial random bit tuple.}}
\label{fig:encas145x157x169}     
\end{figure}

Without pretending here to be exhaustive in surveying the emergent features of the ENCA family,
we conclude this preliminary presentation with the diagrams for ENCA's 145, 157, 169 (Figure \ref{fig:encas145x157x169}). 
Starting from the \emph{same tuple of random bits}, 
all three automata exhibit unbounded name creation, while allowing names to `survive' for considerably long times.

 %
%
%
%
\section{Conclusions}
\label{sect:Conclusions}
%
%
We have only scratched the surface of the attractive topic of NCA particles.
Several questions need to be addressed rigorously. The most prominent ones are:
How many types of NCA particle can we conceive and formally define?
How are they automatically detected?  
Which logical operators best describe them?
What types of computation can they carry out?

The lesson from classical CA's suggests that, beside particles, 
other interesting emergent properties could be observed in ENCA and NCA computations,
most notably fractals and pseudo-randomness.  These features may also require re-definition,
under the nominal perspective.

One of the appealing (and bold) postulates of the computational universe conjecture
is that of a fundamental determinism, meaning that no coin-flipping -- no constant creation of information out of the blue -- 
would be necessary for sustaining the dynamics of the universe, 
in contrast with views attributing a fundamental role to probabilities and stochastic processes.
How does unbounded name creation qualify in this respect? How realistic is it, in physical terms?
Is it correct to equate it to a stochastic model?

In spite of the rich variety of nature-like phenomena that CA's exhibit,
it has been observed that this model has a stronger value 
as a metaphor for the computational universe,
than as an actual, realistic candidate for a fundamental physical theory (of everything).
Then, once these and many other open questions about Nominal Cellular Automata are addressed satisfactorily,
we might be in a better position to transpose the powerful concepts of Nominal Set Theory and
Nominal Computing to other, more attractive and realistic models of spacetime -- notably graph-oriented
models  such as (algorithmic) causal sets.
The importance that the permutation group plays both in nominal sets and in some approaches to the dynamics
of causal sets appears to us as most encouraging for this extended investigation.


\bibliographystyle{eptcs}
\bibliography{all.bib}

\end{document}